\documentclass[prd,aps,twocolumn,a4paper,showkeys,nofootinbib]{revtex4-1}
\usepackage{amsmath}
\usepackage{amsfonts}
\usepackage{amssymb}	
\usepackage{graphicx}
\usepackage{bm}
\usepackage{color}
\usepackage{commath}
\usepackage{xcolor}
\allowdisplaybreaks

\def\GMc2{G M_{\odot} c^{-2}}

\def\lm{{\ell m}}

\def\lm{{\ell m}}

\def\lm{{\ell m}}

\def\l{{\ell }}

\def\F{{\cal F}}

\newcommand\be{\begin{equation}}
\newcommand\ee{\end{equation}}

\def\TEOBResumS{\texttt{TEOBResumS}}
\def\TEOBResumSDali{\texttt{TEOBResumS-DALI}}

\begin{document}
\title{Towards a gravitational self force-informed effective-one-body waveform model\\
for nonprecessing, eccentric, large-mass-ratio inspirals}
\author{Alessandro \surname{Nagar}${}^{1,2}$}
\author{Simone \surname{Albanesi}${}^{1,3}$}
\affiliation{${}^1$INFN Sezione di Torino, Via P. Giuria 1, 10125 Torino, Italy}
\affiliation{${}^2$Institut des Hautes Etudes Scientifiques, 91440 Bures-sur-Yvette, France}
\affiliation{${}^{3}$ Dipartimento di Fisica, Universit\`a di Torino, via P. Giuria 1, 
10125 Torino, Italy}

\begin{abstract}
Building upon several recent advances in the development of effective-one-body models for spin-aligned 
eccentric binaries with individual masses $(m_1,m_2)$  we introduce a new EOB waveform model that aims 
at describing inspiralling binaries  in the large mass-ratio regime, $m_1\gg m_2$.
The model exploits  the current state-of-the-art \TEOBResumSDali{} model for eccentric binaries, 
but the standard EOB potentials $(A,\bar{D},Q)$, informed by Numerical Relativity (NR) simulations, are replaced with 
the corresponding functions that are linear in the symmetric mass ratio $\nu\equiv m_1 m_2/(m_1+m_2)^2$ taken 
at 8.5PN accuracy. To improve their strong-field behavior, these functions are: (i) suitably factorized and resummed
using Pad\'e approximants and (ii) additionally effectively informed to state-of-the-art numerical results obtained by 
gravitational self-force theory (GSF). For simplicity, the spin-sector of the model is taken to be the one of \TEOBResumSDali{}, 
though removing the NR-informed spin-orbit effective corrections. We propose the current GSF-informed EOB framework 
as a conceptually complete analytical tool to generate waveforms for eccentric Extreme (and Intermediate) Mass Ratio Inspirals 
for future gravitational wave detectors.
\end{abstract}
   
\date{\today}

\maketitle

\section{Introduction}
The inspiral of a stellar mass compact object into a massive ($\sim 10^4-10^7M_\odot$)
black hole (BH) generates a complicated gravitational wave (GW) signal that is expected to be one
of the prime source of the space-based GW detector LISA (Laser Interferometer Space Antenna)~\cite{LISA:2017pwj}.
These extreme mass-ratio inspirals (EMRIs) have the potentialities of unveiling and testing deep features
of strong-field General Relativity~\cite{Berry:2019wgg}. To do so, an accurate modeling of the emitted 
waveform is needed, since it has to accurately match the astrophysical signal over the tens of thousands of 
wave cycles that will be in the LISA's band.
The evolution of these large mass ratio binaries has been approached via perturbation theory and 
the computation of the gravitational self-force (GSF) exerted by the field of the small 
black hole on itself~\cite{Pound:2015tma,Barack:2018yvs,Miller:2020bft,Hughes:2021exa,Pound:2021qin,Wardell:2021fyy,Warburton:2021kwk}.
Explicit implementations of full GSF evolution appeared recently~\cite{VanDeMeent:2018cgn,Lynch:2021ogr},
in particular for the case of eccentric orbits around a Kerr black hole.
Despite the recent progresses in the field, a lot remains to be done on the GSF side to 
comprehensively account for the complete phenomenology that is expected for an EMRI.
A generic EMRIs is in fact imagined to be an eccentric binary where {\it both} objects 
are spinning and the spins are not aligned with the orbital angular momentum. More
specifically also the spin of the object with the smaller mass (usually called the secondary)
is important and cannot be neglected~\cite{Piovano:2020zin,Piovano:2021iwv,Mathews:2021rod,Skoupy:2022adh,Timogiannis:2022bks}.

Although an EMRI can be seen as a small perturbation of the Kerr metric, and this is
the essential hypothesis behind any GSF-based calculation, the binary dynamics (and waveform), 
with its full complexity and with {\it all} effects, is naturally described within the effective-one-body (EOB) framework.
This method~\cite{Buonanno:1998gg,Buonanno:2000ef,Damour:2000we,Damour:2001tu,Damour:2015isa} 
is a way to deal with the general-relativistic two-body problem that, by construction, allows the inclusion of 
perturbative (e.g. obtained using post-Newtonian methods or BH perturbation theory) 
and  non perturbative (e.g., full numerical relativity results) within a single theoretical framework. 
Since the extreme mass ratio limit is included in the formalism by construction, the EOB model
is the natural framework to deal with EMRIs~\cite{Yunes:2009ef,Yunes:2010zj}.

For comparable mass binaries, as target sources for ground-based detector, a successful strategy to 
compute highly accurate waveform templates is to use the EOB approach informed by a limited 
amount of NR simulations. One then uses different sets of NR simulations to 
validate the model~\cite{Nagar:2020pcj,Riemenschneider:2021ppj,Nagar:2021xnh,Albertini:2021tbt}.
Analogously, for EMRIs one can construct an EOB model {\it informed} by exact GSF calculation~\cite{Damour:2009sm,Barack:2010ny}.
The same model could then be validated against full GSF evolutions, 
that have become available very recently~\cite{Wardell:2021fyy,Warburton:2021kwk}.
In particular, a step towards incorporating full 1GSF information (i.e., linear in $\nu$) in the Hamiltonian 
was done by Antonelli et al.~\cite{Antonelli:2019fmq}, building upon previous work~\cite{Barausse:2011dq,LeTiec:2011dp}.
They used the post-Schwarzschild Hamiltonian~\cite{Damour:2017zjx} in the energy 
gauge~\cite{Bini:2020wpo} so to overcome the well-known problems related to 
the presence of the light-ring coordinate singularity in the standard EOB gauge 
(or Damour-Jaranowski-Sch\"afer, DJS hereafter~\cite{Damour:2000we})~\cite{Akcay:2012ea}. 
Although promising, the approach  of~\cite{Antonelli:2019fmq}, that was limited to the case of 
nonspinning binaries, needs more development to construct a complete model, informed by Numerical 
Relativity simulations, able to span the full range of mass ratios.
By contrast, the use of {\it dissipative} self-force results in EOB model is a crucial element
in the construction of a highly accurate waveform and radiation reaction 
force~\cite{Damour:2008gu,Pan:2010hz,Nagar:2011aa,Taracchini:2013wfa,Nagar:2016ayt,Messina:2018ghh,Albanesi:2021rby,Albanesi:2022ywx}.

Our main interest here is to introduce a  GSF-informed EOB waveform model for EMRIs.
Since the gravitational signals produced by these systems can remain in the sensitivity band of 
LISA for months or even years, the contribution to the total SNR of the merger-ringdown signal is negligible. 
For this reason, we focus here on the inspiral of the system and adopt a different strategy from 
Ref.~\cite{Antonelli:2019fmq}. We build upon \TEOBResumSDali{}, but we incorporate GSF 
information in the EOB metric potentials expressed in the  DJS gauge. The dissipative contributions to
the dynamics that we employ are the standard ones of \TEOBResumSDali{} and we leave 
further developments in this direction to future work.

The paper is organized as follows. In Sec.~\ref{sec:eob} we give a brief recap 
of the EOB model, in particular discussing the resummed and GSF-informed EOB potentials.
In Sec.~\ref{sec:example} we provide a case example for the waveform,
while Sec.~\ref{sec:end} collects our conclusions and the steps to be
undertaken in the future. We use geometrized units with $G=c=1$.

\section{Effective-one-body dynamics and waveform with GSF inputs}
\label{sec:eob}
The structure of the dynamics of the GSF-informed model we are going to
discuss here is the same as Ref.~\cite{Nagar:2021gss} except for the structure
of the EOB potentials $(A,D,Q)$ and their resummed representation. Before diving
in the details, let us recall the basic notation adopted.
We use mass-reduced phase-space variables $(r,\varphi,p_\varphi,p_{r_*})$,  related to the physical 
ones by $r=R/M$ (relative separation), $p_{r_*}=P_{R_*}/\mu$ (the tortoise-coordinate radial momentum), 
$\varphi$ (orbital phase), $p_\varphi=P_\varphi/(\mu M)$ (angular momentum) and $t=T/M$ (time),
where $\mu\equiv m_1 m_2/M$ and $M\equiv m_1+m_2$. 
The tortoise-coordinate radial momentum is related to the radial momentum, conjugate to $r$, as
$p_{r_*}\equiv (A/B)^{1/2}p_r$, where $A$ and $B$ are the EOB potentials 
(with included spin-spin interactions~\cite{Damour:2014sva}) 
and $D=A B$ (for nonspinning systems).
The EOB Hamiltonian is $\hat{H}_{\rm EOB}\equiv H_{\rm EOB}/\mu=\nu^{-1}\sqrt{1+2\nu(\hat{H}_{\rm eff}-1)}$, 
with $\nu\equiv \mu/M$ and $\hat{H}_{\rm eff}=\tilde{G}p_\varphi + \hat{H}^{\rm orb}_{\rm eff}$, 
where $\tilde{G}p_\varphi$ incorporates odd-in-spin (spin-orbit) effects while 
$\hat{H}^{\rm orb}_{\rm eff}$ takes into account even-in-spin effects through the use of 
the centrifugal radius $r_c$~\cite{Damour:2014sva}. The orbital Hamiltonian for 
non-spinning systems reads
\be
\hat{H}_{\rm orb}=\sqrt{A(u)\Big((1+p_\varphi u^2)  + Q(u,p_{r_*};\nu)\Big)+ p_{r_*}^2},
\ee
where $u\equiv 1/r$ and $Q$ is the generalized mass-shell function. 
The spin sector we use here is  the same as~\cite{Nagar:2021gss}, although we set to 
zero the NR-informed next-to-next-to-next-to-leading order effective 
coefficient $c_3$. This choice for the spin-orbit sector is, at the moment, merely illustrative,
since there is a much wider amount of analytical information recently obtained that could 
be used, see e.g.~\cite{Antonelli:2020aeb,Antonelli:2020ybz}. From now on we focus 
predominantly on the nonspinning limit of the model and discuss in detail the structure of 
the potentials with 1GSF information. These potentials can then be dressed by the spin-dependent
factor so to incorporate spin-spin interaction in the usual way using the centrifugal radius~\cite{Damour:2014sva}.

\subsection{GSF-informed potentials}
\label{sec:gsf}
At linear order in $\nu$, the effective EOB potentials can be written as
\begin{align}
A(u;\nu) & = 1-2 u + \nu a_{\rm 1SF}(u) + {\cal O}(\nu^2),\\
\bar{D}(u;\nu) &= \dfrac{1}{AB}= 1 + \nu \bar{d}_{\rm 1SF}(u) + {\cal O}(\nu^2),\\
Q(u,p_{r_*};\nu) &= \nu q_{\rm 1SF}(u) p_{r_*}^4 .
\end{align}
The functions $(a_{\rm 1SF}(u),\bar{d}_{\rm 1SF}(u),q_{\rm 1SF}(u))$ are taken
 here as PN expansions, although they have been computed numerically~\cite{Akcay:2012ea,Akcay:2013wfa,vandeMeent:2015lxa,Akcay:2015pjz} from self-force results. 
 Using the numerically computed values in the model would require some sort
of accurate interpolation or fits, like those performed in Ref.~\cite{Akcay:2012ea} for
$a_{\rm 1SF}(u)$ or for $\bar{d}$ and $q_{\rm 1SF}$ in Ref.~\cite{Akcay:2015pjz}. In particular, we will refer,
and use, the model fit $\# 14$ of~\cite{Akcay:2012ea} as exact representation of $a_{\rm 1SF}$,
while for $(\bar{d}_{\rm 1SF},q_{\rm 1SF})$ we directly use the numerical points obtained
in Ref.~\cite{Akcay:2015pjz}.
The purpose of this work is to rely as much as possible on analytical results for
the functions $(a_{\rm 1SF},\bar{d}_{\rm 1SF},q_{\rm 1SF})$, i.e. on truncated PN
series that will need some additional resummation to improve their behavior in
the strong-field regime. In doing so, we aim at reducing to the minimum the use
of exact GSF results obtained numerically. More precisely,  we want to use them 
only to {\it improve}, possibly in simple ways, the performance of the resummed PN series,
as we will see below.
We will mostly rely on the 8.5PN-accurate results for the three functions obtained 
in Ref.~\cite{Bini:2014nfa}. One of the important take away messages of Ref.~\cite{Bini:2014nfa}
was that, despite the factorization of the light-ring pole singularity, the residual PN-series 
of $a_{\rm 1SF}$ oscillates in strong field. This makes the, still valuable, 
analytical information not useful for waveform modelization purposes.
This situation is somehow reminiscent of what found in Ref.~\cite{Damour:1997ub}
when analyzing the PN-expanded energy flux of a test-mass on circular orbits around
Schwarzschild. The flux is singular at the Schwarzschild light-ring (as is the case of $a_{\rm 1SF}$)
but then, once the pole singularity is factored out, it is possible to robustly resum the
residual functions via Pad\'e approximants~\cite{Damour:1997ub}. 
In addition, Ref.~\cite{Damour:2007yf} also showed that it is possible to improve 
the accuracy of such Pad\'e resummations by tuning an effective parameter to the exact 
data of the flux for circular orbits (see Fig.~1 of~\cite{Damour:2007yf}).
We proceed here following in spirit Refs.~\cite{Damour:1997ub,Damour:2007yf},
applying the same rationale to $a_{\rm 1SF}$ and to the other two potentials as well.
In more detail: (i) we apply a certain factorization and resummation procedure, 
to be discussed below, to improve and stabilize the behavior of the truncated PN 
series in strong field, and then (ii) we additionally fit to the GSF-exact numerical 
data a correcting factor to improve them up to (and below) the LSO.
This will give us access to relatively simple, but GSF-informed, analytical representations 
of $(a_{\rm 1SF},\bar{d}_{\rm 1SF},q_{\rm 1SF})$ that are robust and accurate in 
strong field. Let us now analyze each EOB potential separately.

\subsubsection{Resumming $a_{\rm 1SF}(u)$}
As pointed out in Ref.~\cite{Akcay:2012ea}, when expressed in the original, standard,
DJS gauge~\cite{Damour:2000we}, the $a_{\rm 1SF}$ function exhibits a coordinate 
singularity at the light-ring (LR). However, as long as we are not interested in the
transition from inspiral to plunge and merger, this has no impact and we can work 
with this gauge\footnote{See Ref.~\cite{Antonelli:2019fmq}
for a different strategy that incorporates GSF information in a different gauge.}. 
Following Ref.~\cite{Bini:2014nfa}, the LR singularity can be factored out, so
to work with the doubly rescaled potential
\be
\hat{a}^E_{\rm 1SF}(u) = \dfrac{a_{\rm 1SF}}{2 u^3 E(u)}\ ,
\ee
where $E(u) = (1-2u)/\sqrt{1-3u}$. In Ref.~\cite{Bini:2014nfa} it was 
noted that, despite this factorization, $\hat{a}^E_{\rm 1SF}(u)$ as truncated 
PN oscillates around the exact numerical values, as represented  by the accurate 
fit $\#14$ of Ref.~\cite{Akcay:2012ea}, see in particular Fig.~1 of Ref.~\cite{Bini:2014nfa}. 
Given the large span by the various PN truncations of $\hat{a}^E_{\rm 1SF}$, 
it seems that their analytical knowledge  is not useful practically.
More recently Ref.~\cite{Kavanagh:2015lva} developed improved techniques to push
the PN-expansion of gauge-invariant quantities at very high PN order, but this does not
seem to solve the problem, as we will see below.
To overcome the difficulty related to the plain PN expansion of $\hat{a}^E_{\rm 1SF}$, 
we present here a new resummation strategy that is loosely inspired to the factorization 
and resummation procedure developed for the PN waveform~\cite{Damour:2008gu}.
When one inspects the PN series of $\hat{a}^E_{\rm 1SF}(u)$ one sees that there are integer powers
and semi-integer powers of $u$. The semi-integer powers of $u$ are related to the hereditary
effects (the tails) that show up in the conservative part of the dynamics after a certain PN
order. In the waveform, the leading-order hereditary corrections are all packed together
in the tail factor~\cite{Damour:2007xr,Damour:2008gu}. The factorization and resummation 
scheme we exploit here is analogous, conceptually, to the one implemented for the 
waveform~\cite{Damour:2008gu}, especially considering that the singularity at the light ring 
has already been factored out above.
The idea is to write the function as the product of two factors: one that only contains integer 
powers of $u$ and the other one that only contains semi-integer powers of $u$. Each factor is then 
resummed separately. In precise terms, the procedure is as follows:
\begin{itemize}
\item[(i)]We start for a given PN-expanded expression of $\hat{a}^E_{\rm 1SF}(u)$, we identify the series in integer
             powers of $u$, we factor it out and re-expand what remains. In practice we have the following
             factorization
             \be
             \hat{a}^E_{\rm 1SF}(u) = \hat{a}_{\rm 1SF}^{E,{\rm integer}}(u)\hat{a}_{\rm 1SF}^{E,{\rm half}}(u^{1/2}) \ ,
             \ee
             where  we  have spelled out explicitly that $\hat{a}_{\rm 1SF}^{E, {\rm integer}}$ 
             is function of powers of $u$, while $\hat{a}_{\rm 1SF}^{E,{\rm half}}(u^{1/2})$ only 
             of odd powers of $u^{1/2}$. We recall that semi-integer powers enter the function at 5.5PN 
             order, so that the factorized functions begin as
             \begin{align}
             &\hat{a}_{\rm 1SF}^{E,{\rm integer}}(u) = 1 + \left(\dfrac{97}{6}-\dfrac{41}{64}\pi^2\right)u\nonumber\\
                                                     & + \left[-\dfrac{138}{5}+\dfrac{1947}{1024}\pi^2+\dfrac{64}{5}\left(\gamma+2\log 2+\dfrac{1}{2}\log u\right)\right]u^2 \nonumber\\
                                                     &+ \bigg(-\dfrac{2269007}{6300}-\dfrac{16309}{210}+\dfrac{63421}{1536}\pi^2\nonumber\\
                                                     &-\dfrac{32693}{210}\log2-\dfrac{16309}{420}\log u\bigg)u^3 +O(u^4) \ ,  \\
             &\hat{a}_{\rm 1SF}^{E,{\rm half}}(u^{1/2}) = 1 + \dfrac{6848}{525}\pi u^{7/2} \nonumber\\
             &+ \pi\left(-\dfrac{6043487}{22050} + \dfrac{4387}{525}\pi^2\right)u^{9/2}\nonumber\\
             &+\pi\Bigg[\dfrac{311719009727}{65488500}
                -\dfrac{473423711}{1411200}\pi^2 + \dfrac{179867}{33600}\pi^4\\
               &-\dfrac{438272}{2625}\left(\gamma +2\log 2+\dfrac{1}{2}\log u \right)\Bigg]u^{11/2}+O(u^{13/2}) \nonumber\ .
             \end{align}
 In the following we will mostly use the variable $v\equiv u^{1/2}$ to express the half-integer factors  and the
 reader should be reminded that the PN truncations of  $\hat{a}_{\rm 1SF}^{E,{\rm half}}$ only involve odd powers of $v$.
\item[(ii)] Each factor is then resummed using Pad\'e approximants. We explored various Pad\'e approximants
(starting from the diagonal ones) and concluded that there is not a standard rule that holds in general. We found
that when the approximants get closer to the diagonal there might be spurious poles. In practice, for each PN
order we have to look for a suitable combination of Pad\'e approximants that allows to construct a function that 
is qualitatively and quantitatively consistency with the exact one.
\end{itemize}
\begin{table}[t]
\begin{center}
\begin{ruledtabular}
\begin{tabular}{c  c c}
PN & $P^n_d\left(\hat{a}_{\rm 1SF}^{E,{\rm integer}}(u)\right)$ & $P^n_d\left(\hat{a}_{\rm 1SF}^{E, {\rm half}}(v)\right)$  \\
\hline
\hline
6 & (2,2)& (7,0)   \\
7 & (3,2)&(8,0) \\
8 & (4,2)& (8,3) \\
9 & (5,2)& (11,2) \\
10 & (5,3)& (10,5)\\
11 & (7,2)& (11,6)\\
12 & (8,2)& (15,4)\\
13 & (7,4)& (16,5)\\
14 & (9,3)& $(19,4)$ \\
15 & (10,3)& (20,5)\\
16 & (11,3)& (21,6)\\
17 & (11,4)& (18,11)\\
18 & (13,3)& (23,10)\\
19 & (13,4)& (23,9)\\
20 & (14,4)& (25,10)\\
21 & (15,4)& (27,10)\\
22 & (16,4)& (27,12)\\
23 & (17,4)& (31,10)\\
24 & (19,3)& (33,10)\\
25 & (20,3) & (33,12) \\
\end{tabular}
\end{ruledtabular}
\end{center} 
\caption{\label{tab:25PN}Best choice of Pad\'e approximants, $P^n_d$, 
(as shown in the middle panel of Fig.~\ref{fig:A25})  for each (integer) PN order 
considered up to 25PN. Note that the the order $(n,d)$ of the Pad\'e approximant 
refers to powers of $u$ for $\hat{a}_{\rm 1SF}^{E, {\rm integer}}$ while 
to powers of $v\equiv u^{1/2}$ for $\hat{a}_{\rm 1SF}^{E, {\rm half}}$.}
\end{table}
\begin{figure}[t]
\includegraphics[width=0.43\textwidth]{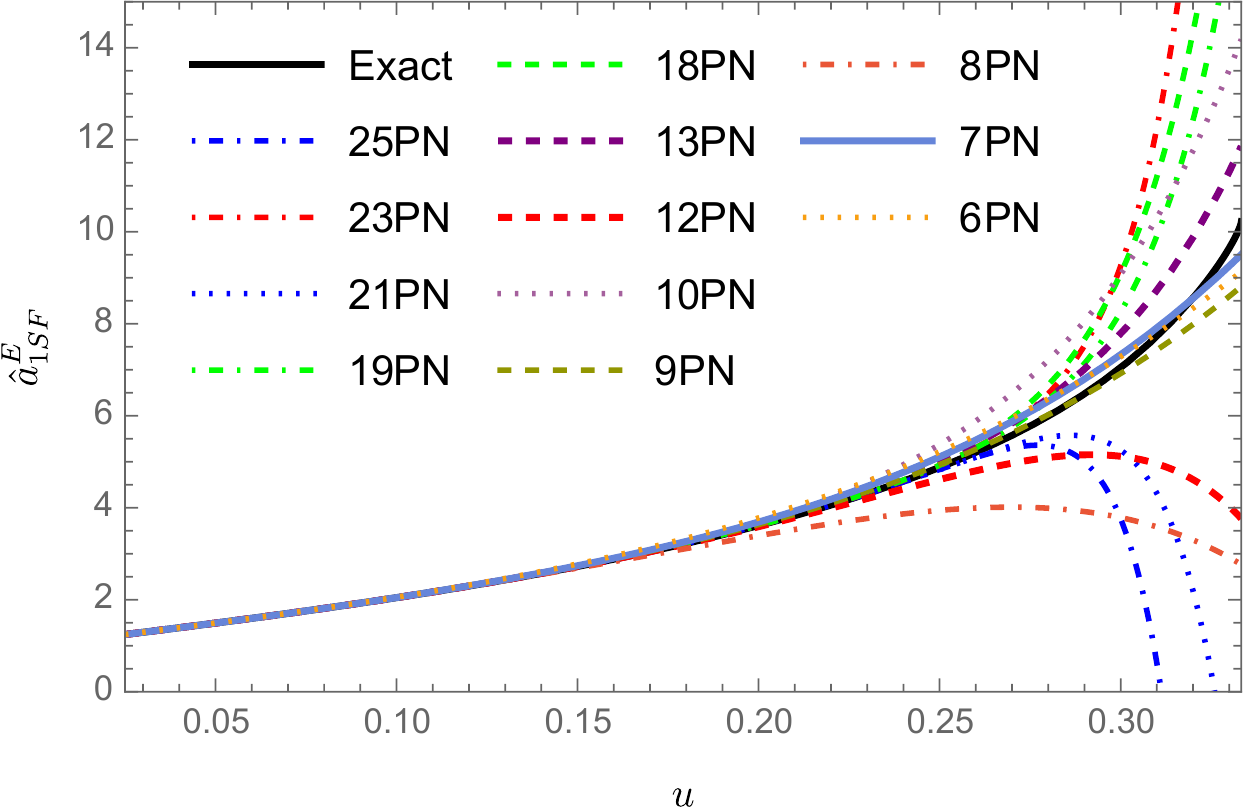}\\
\vspace{2.5 mm}
 \includegraphics[width=0.43\textwidth]{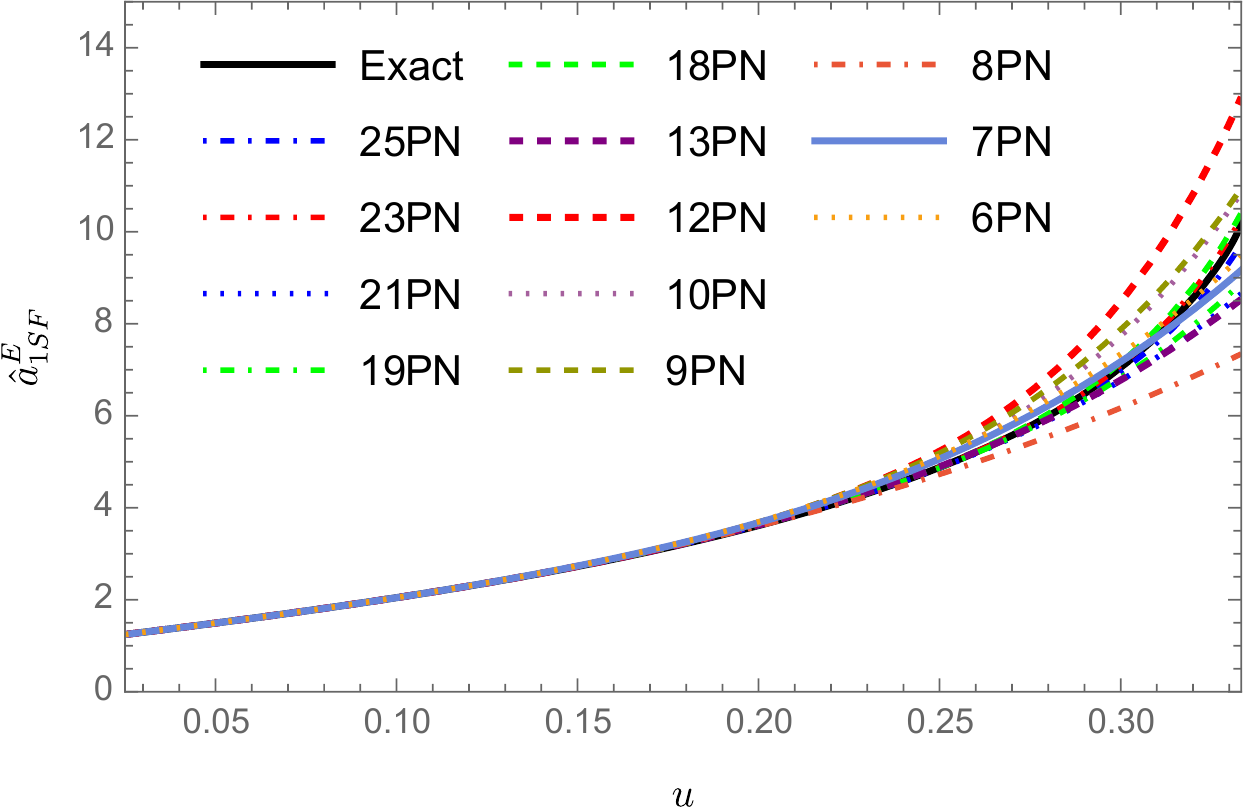}\\
 \vspace{2.5mm}
 \includegraphics[width=0.43\textwidth]{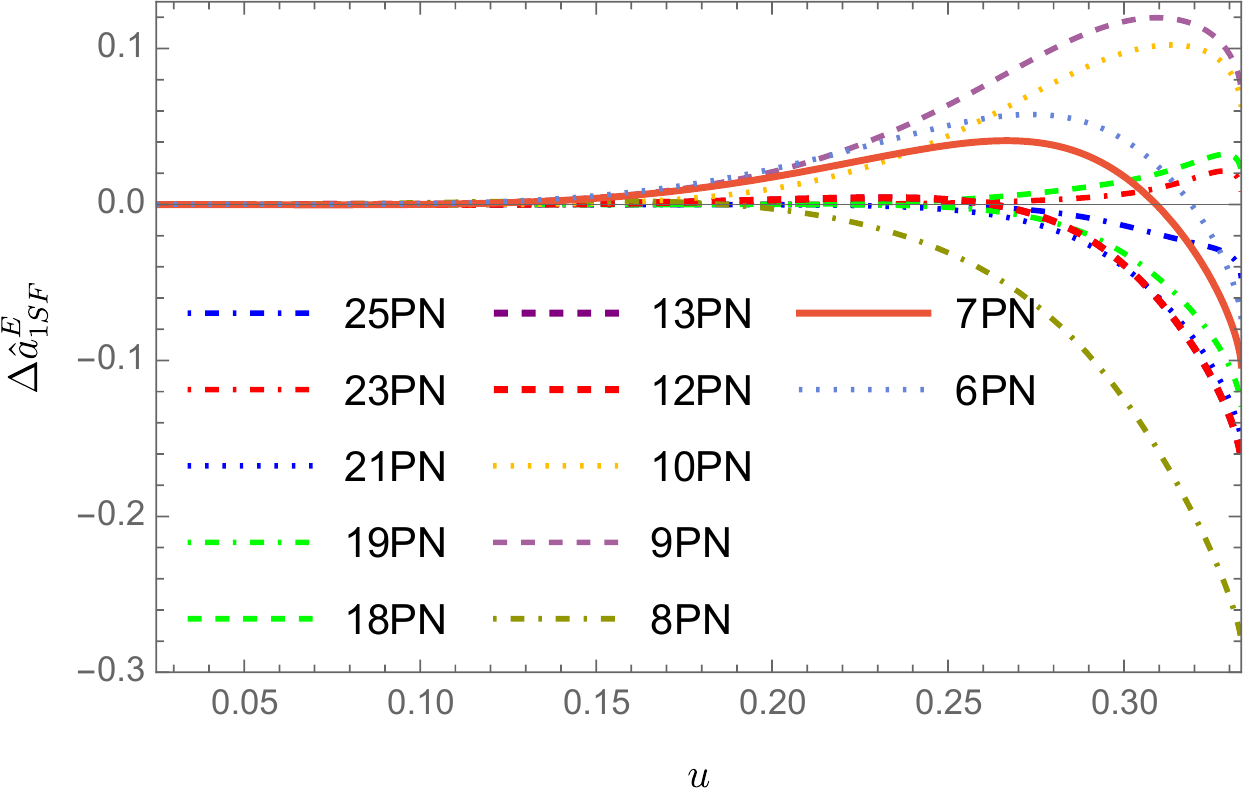}
\caption{\label{fig:A25}Exploring the behavior of the PN-expansion of $\hat{a}^{E}_{\rm 1SF}$ up to
25PN accuracy. Top panel: the Taylor-expanded function for some selected, PN orders. The various
PN truncations randomly oscillate around the exact function with large variations between one order
and the other. Middle panel: the factorized and resummed functions, taking the Pad\'e approximant
of each factor as indicated in Table~\ref{tab:25PN}. Bottom panel: relative differences of the resummed functions with the exact 
results. The 25PN resummed is consistent with exact data at $\sim 4\%$ near the light ring.}
\end{figure}
%
\begin{figure}[t]
\hspace{0.57cm}
\includegraphics[width=0.42\textwidth]{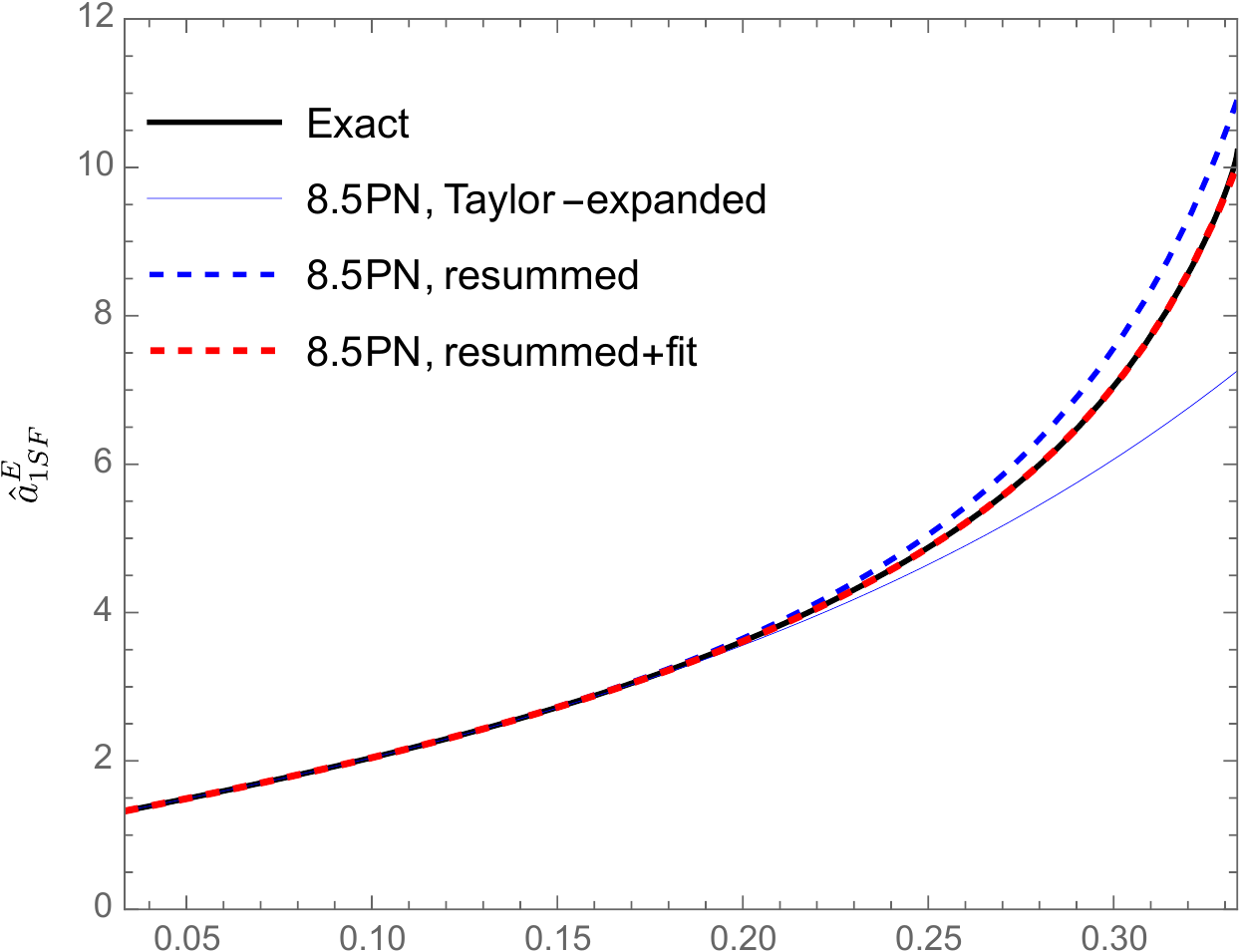}\\
\vspace{4mm}
\includegraphics[width=0.46\textwidth]{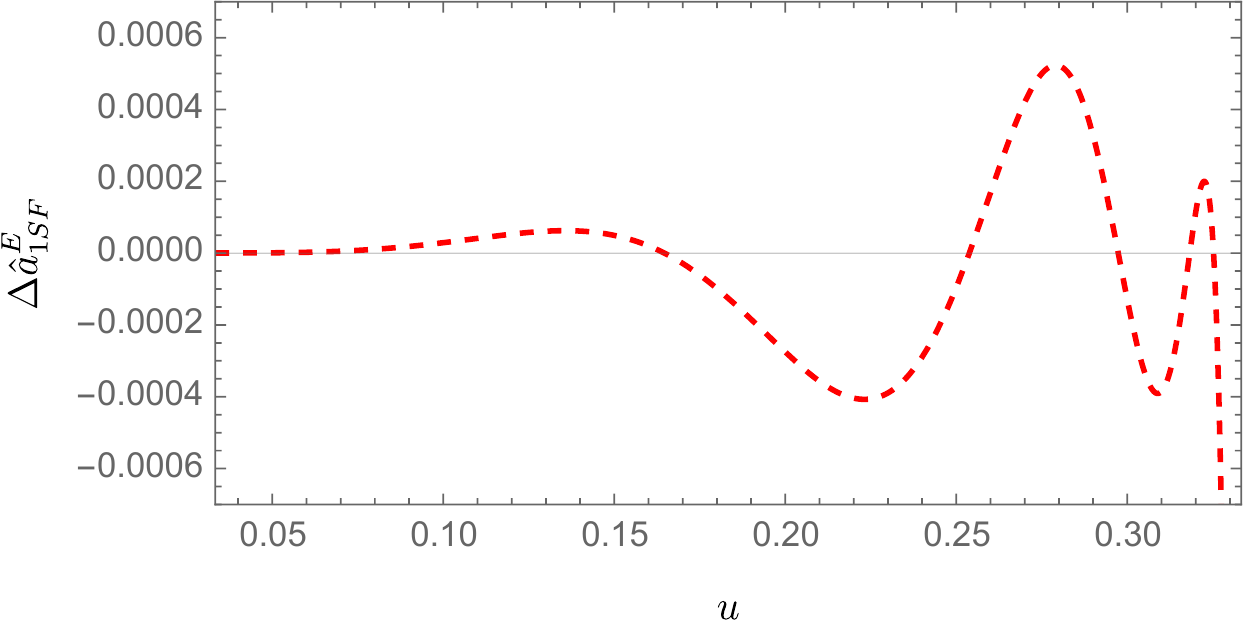}
\caption{\label{fig:A}Top panel: the exact $\hat{a}^E_{\rm 1SF}$ from Ref.~\cite{Akcay:2012ea} is 
compared with the 8.5PN-expanded one, with its resummed homologous and with the one with 
the GSF-informed effective correction $f_{a_{\rm 1SF}}(u)$ of Eq.~\eqref{eq:fau}. 
Bottom panel: the relative difference between this latter and the exact function.
}
\end{figure}
\begin{table}[t]
\begin{center}
\begin{ruledtabular}
\begin{tabular}{l  c c c c}
model & $u_{\rm LSO}(10^{-3})$ & $u_{\rm LSO}(10^{-4})$ & $u_{\rm LSO}(10^{-6})$ & \\
\hline
\hline
Exact &0.1668169583 & 0.1666816489 &  0.1666668164\\
 8.5PN &   0.1668127576      &0.1666812329 & 0.1666668123\\
 8.5PN$_{\rm resummed}$   & 0.1668202025& 0.1666819700& 0.1666668196 \\
 8.5PN$_{\rm resummed}^{\rm GSF_{\rm tuned}}$ & 0.1668170574 & 0.1666816588 & 0.1666668165\\
\end{tabular}
\end{ruledtabular}
\end{center} 
\caption{\label{tab:8p5PN}Location of the LSO for various 8.5PN-accurate approximants contrasted with
to the exact values (top row). The PN-expanded expression is {\it locally} better than the resummed one,
that is however crucially improved by the GSF-informed correcting factor.}
\end{table}
Table~\ref{tab:25PN} reports the choice of Pad\'e approximants for all PN orders we considered, from 
6PN to 25PN. The performance of a selection of PN orders is illustrated in Fig.~\ref{fig:A25}.
The top panel of the figure reports a selection of the simple, Taylor-expanded, functions at various
PN orders, and it is the analogous of Fig.~1 of Ref.~\cite{Bini:2014nfa}. One sees that, 
despite the presence of high PN orders, there is no evidence of convergence to the exact
function (black line). This latter is taken to be fit $\#14$ of Ref.~\cite{Akcay:2012ea}, that
is equivalent to the real numerical data for any practical purpose. However, applying the factorization 
and resummation introduced above, with the selected Pad\'e approximants of Table~\ref{tab:25PN}, 
it is possible to somehow stabilize the PN result and to obtain a more consistent behavior of the various 
PN orders, either among themselves and with the exact function, as evident from the middle and bottom 
panel of Fig.~\ref{fig:A25}. One can note in particular the excellent numerical/analytical agreement
found for 23PN or 25PN, with fractional differences at the LSO that are $\sim 4.3259\times 10^{-6}$
and $-7.2959\times 10^{-7}$ respectively. These value worsen to $\simeq -2.1\%$ and $\simeq -4\%$ 
close to the light ring respectively.
Although this certainly marks a progress in the best use of PN results, it is still not very practical to
use such large analytical expression in numerical codes. In addition, there are still differences that
are nonnegible towards the light ring. The plot however illustrates that the factorization and 
resummation procedure is very effective and makes thigh-PN (resummed) expressions 
somehow redundant among themselves. As a simplifying strategy that also takes advantage
of the exact GSF information known numerically, we follow a different strategy: (i) we choose 
a relatively low and manageable PN order that, once resummed exhibits a reasonable level 
of agreement with the exact function; (ii) we slightly modify the resulting, resummed, analytical 
function with an effective corrective factor that is fitted to the exact data for improved improved accuracy.
To pursue this strategy, from now on we only work at 8.5PN. This gives an acceptable compromise 
between analytical simplicity and accuracy of the factorized and resummed expression. 
The resummation procedure yields 
\be
\label{eq:a1SFtot}
\hat{a}_{\rm 1SF}^{\rm E,8.5PN}=P^3_3\left(\hat{a}^{E,{\rm integer}}_{\rm 1SF}\right)P^7_6\left(\hat{a}_{\rm 1SF}^{E,{\rm half}}(v)\right)f_{a_{\rm 1SF}}(u) \ ,
\ee
where we have also introduced $f_{a_{\rm 1SF}}(u)$ as a correcting factor to be fitted to the numerical data. 
It is chosen to be
\begin{align}
\label{eq:fau}
f_{a_{\rm 1SF}}(u) &= 1 + \left\{c_7 + c_7^{\rm log}\log u +c_7^{\log^2}\log ^2\!u\right\}u^7 \ .
\end{align}
Figure~\ref{fig:A} shows together the exact  $\hat{a}_{\rm 1SF}^{\rm E}$ function, the 8.5PN accurate  one, 
the resummed one and the GSF-informed one with this correcting function.
The fit is done extracting a list of points in the interval $\Delta u= [1/150,1/3.1]$ and computing $a_{\rm 1SF}(u)$ using the fit model 14 of Ref.~\cite{Akcay:2012ea}.
The values of the fitting coefficients are
\begin{align}
c_7 & = 963.4329,\\
c_7^{\log} &= 811.3827,\\
c_7^{\log^2} & = -205.2558.
\end{align}
The bottom panel of Fig.~\ref{fig:A} shows the fractional difference between Eq.~\eqref{eq:a1SFtot} and the 
exact curve up to the light-ring. At a more quantitative level, it is also interesting to compare the estimate of 
the location of the Last Stable Orbit (LSO) using the various functions of Fig.~\ref{fig:A}. This is reported in
Table~\ref{tab:8p5PN} for three values of the symmetric mass ratio $\nu=(10^{-3},10^{-4},10^{-6})$. It is clear
that, especially for the case of a standard EMRI, our novel use of analytical results offers an excellent
representation of the exact data.
\begin{figure}
\vspace{0.15cm}
\hspace{0.65cm}
\includegraphics[width=0.424\textwidth]{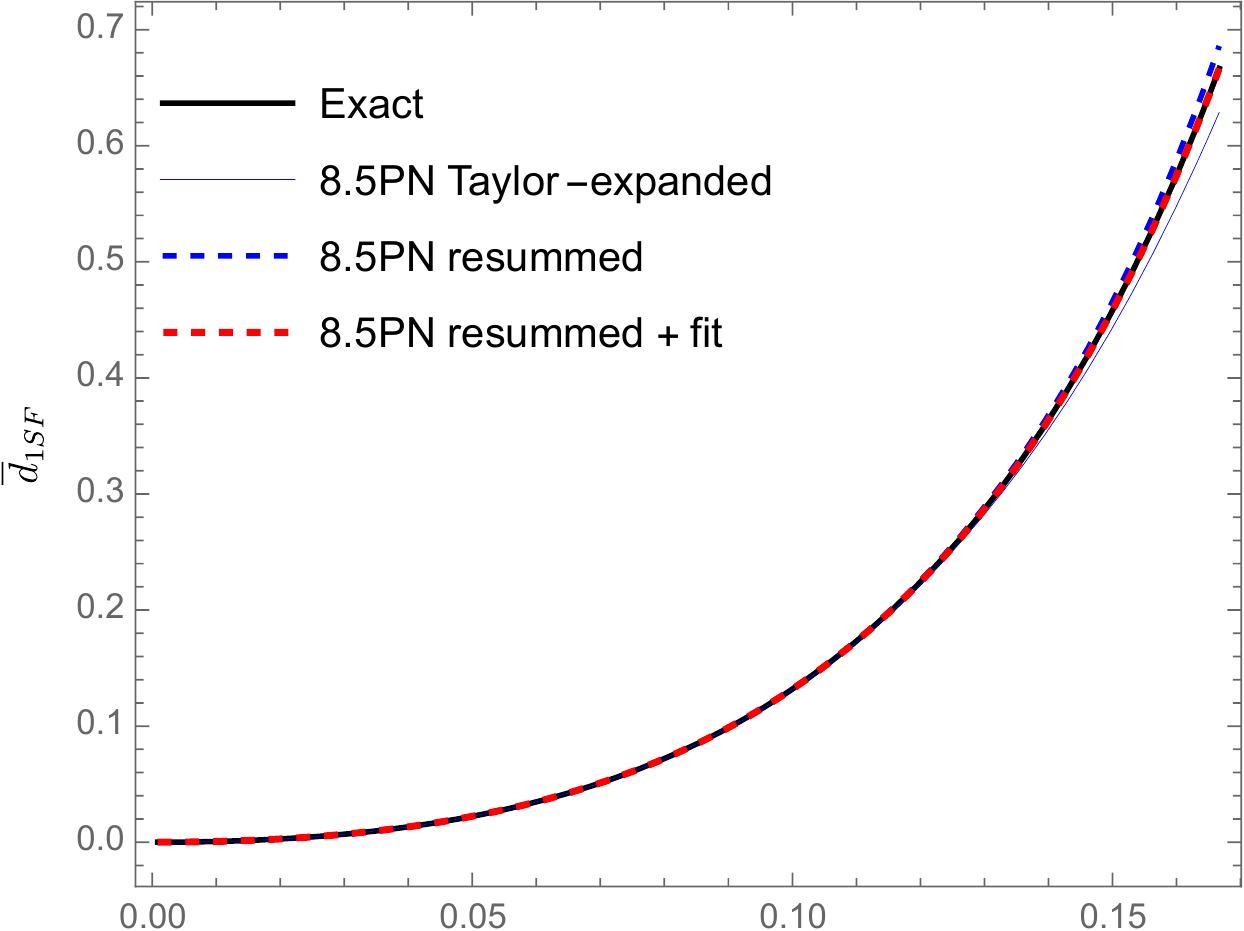}\\
\vspace{2mm}
\hspace{2mm}
\includegraphics[width=0.45\textwidth]{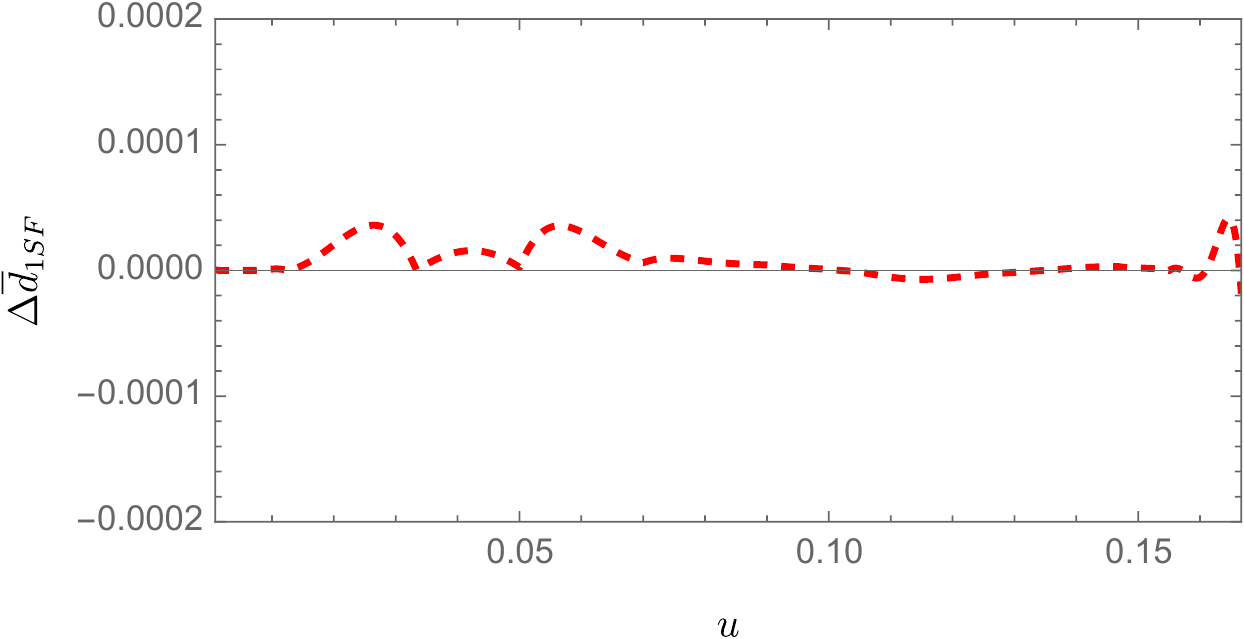}
\caption{\label{fig:d}Top panel: the "exact" $\bar{d}_{\rm 1SF}$ is compared with the 8.5PN-expanded one,
with its resummed homologous and with the one with the additional effective correction. Bottom panel: relative difference
between the GSF-corrected function and the numerical data.}
\end{figure}

\subsubsection{Resumming $\bar{d}_{\rm 1SF}(u)$}
We follow an analogous procedure for the function $\bar{d}_{\rm 1SF}(u)$. The function is given
as a truncated PN series up to $u^{8.5}$ that reads
\be
\label{eq:d1SF}
\bar{d}_{\rm 1SF} = d_2 u^2 + d_3 u^3 +\dots + d_{8.5}u^{17/2} , \
\ee
so that we work with the function $\hat{d}_{\rm 1SF}\equiv \bar{d}_{\rm 1SF}/(d_2 u^2)$.
As above, the function is factorized in a part with integer powers and a part with only
odd powers of $v$. Each factor is Pad\'e resummed so to get
\be
\label{eq:hatd}
\hat{d}_{\rm 1SF}^{\rm resummed}=P^3_3\left(\hat{d}_{\rm 1SF}^{\rm integer}\right)P^7_6\left(\hat{d}^{\rm half}_{\rm 1SF}(v)\right)f_{d_{\rm 1SF}}(u),
\ee
where $f_{d_{\rm 1SF}}$ is the effective correction that is GSF-informed to exact GSF data analogously to $f_{a_{\rm 1SF}}$.
In this case, the fit is performed on the numerical data shown in Table III of Ref.~\cite{Akcay:2015pjz}.
Note that here the numerical results are computed only up
to the LSO since they have been obtained perturbing moderately eccentric orbits, while the computation 
of $d_{\rm 1SF}(u)$ beyond the LSO would
require orbits with higher eccentricity, and even hyperbolic configurations 
for radii close to the LR (see e.g. Ref.~\cite{Barack:2019agd}).
The fit is chosen to incorporate corrections with powers higher than $u^{13/2}$ and is given by
\begin{align}
f_{d_{\rm 1SF}}(u) &= 1 + \left[d_7 + d_7^{\log}\log(u)+d_7^{\log^2}\log^2\!u\right]u^7 \nonumber \\
                             & + \left[d_{7.5}+d_{7.5}^{\log}\log u\right]u^{15/2}\ ,
\end{align}
where the fit coefficients are determined to be
\begin{align}
d_7 &= 10018073.1897\ ,\\
d_7^{\log}&=2634075.0796 \ ,\\
d_7^{\log^2}&=198577.8047 \ ,\\
d_{7.5} & = -9898639.6611 \ ,\\
d_{7.5}^{\log}&=2601282.0285 \ .
\end{align}
The quality of the various analytical approximation is illustrated in Fig.~\ref{fig:d}. The exact function is 
contrasted with: (i) the plain 8.5PN-accurate function as calculated in Ref.~\cite{Bini:2014nfa}; 
(ii) its resummed version without the GSF-informed correction factor; (iii) the full, GSF-informed, 
function obtained from Eq.~\eqref{eq:hatd}. The bottom panel of the figure displays the relative
difference between the exact and the GSF-informed function.

\subsubsection{Resumming $q_{\rm 1SF}(u)$}
\begin{figure}[t]
\hspace{0.42cm}
\includegraphics[width=0.422\textwidth]{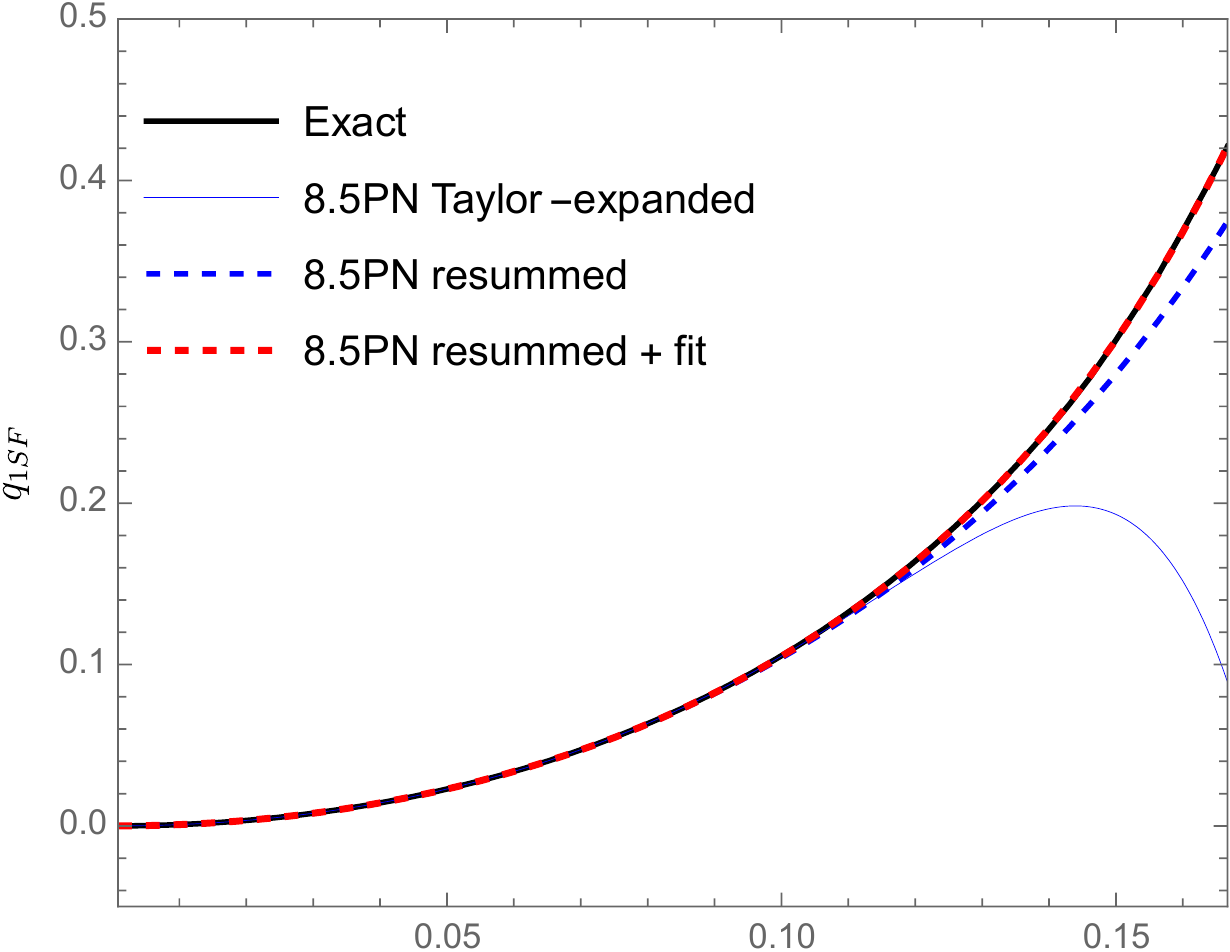}\\
\vspace{2mm}
\includegraphics[width=0.45\textwidth]{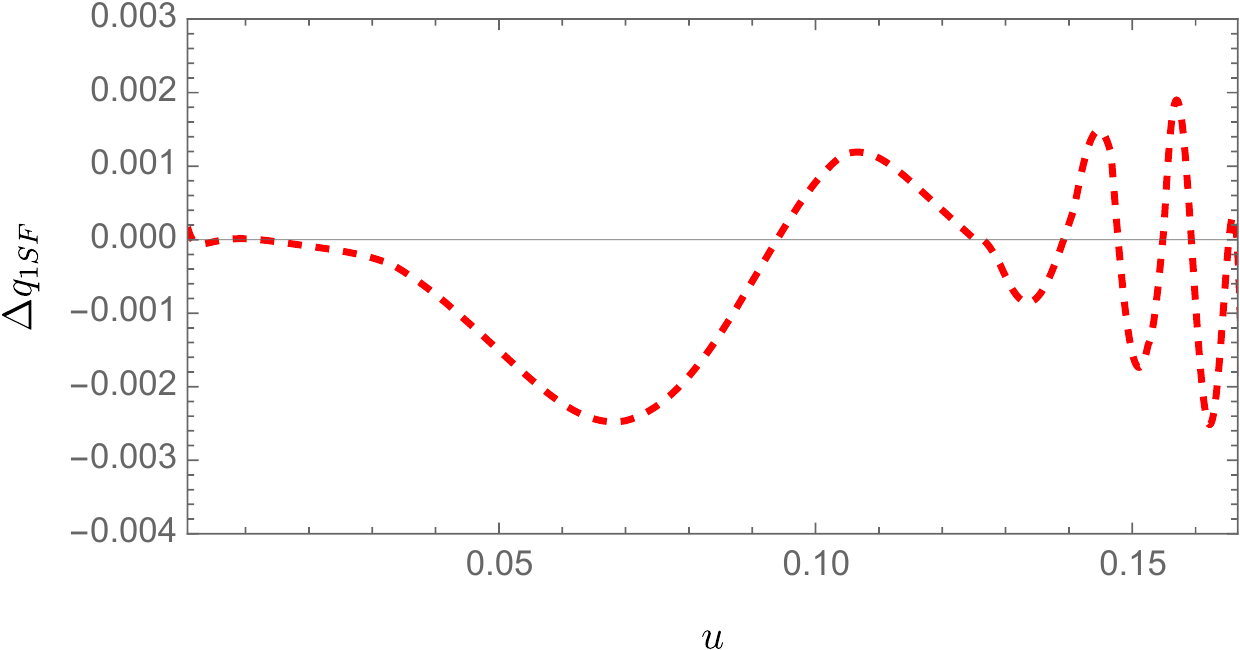}
\caption{\label{fig:q}Top panel: the "exact" $\hat{q}_{\rm 1SF}$ is compared with the 8.5PN-expanded one,
with its resummed homologous and with the one with the additional effective correction. Bottom panel: relative difference
between the GSF-corrected function and the numerical data.}
\end{figure}
Let us finally move to the $q_{\rm 1SF}$ function. We also take this at 8.5PN accuracy, so that it formally reads
\be
q_{\rm 1SF}=q_2 u^2 + q_3 u^3 + \dots + q_{8.5}u^{17/2} \ ,
\ee
then we define $\hat{q}_{\rm 1SF}\equiv q_{\rm 1SF}/(q_2 u^2)$. As above, we factor out the contribution with
integer powers and the one with semi-integer powers. Each one is then resummed with the following Pad\'e approximants
\be
\label{eq:hatq}
\hat{q}_{\rm 1SF}^{\rm resummed}=P^3_3\left(\hat{q}_{\rm 1SF}^{\rm integer}\right)P^7_6\left(\hat{q}^{\rm half}(v)\right)f_{q_{\rm 1SF}}(u),
\ee
while $f_{q_{\rm 1SF}}(u)$ is the correcting function. It is chosen to have the same analytical structure of $f_{d_{\rm 1SF}}$
as
\begin{align}
f_{q_{\rm 1SF}}(u) &= 1 + \left[q_7 + q_7^{\log}\log(u)+q_7^{\log^2}(\log(u))^2\right]u^7 \nonumber \\
                             & + \left[q_{7.5}+q_{7.5}^{\log}\log(u)\right]u^{15/2}\ ,
\end{align}
and the fitting coefficients read
\begin{align}
q_7 &= -6555554285.3947\ ,\\
q_7^{\log}&= -1775629690.3454 \ ,\\
q_7^{\log^2}&=-146188670.5646\ ,\\
q_{7.5} & = 6473498950.5603 \ ,\\
q_{7.5}^{\log}&= -1641347218.6108\ .
\end{align}
Similarly to the $d_{\rm 1SF}(u)$ case, the fit is 
done using numerical GSF data up to the LSO. The numerical 
results, obtained building upon Ref.~\cite{LeTiec:2015kgg}, can be found in Table III of Ref.~\cite{Akcay:2015pjz}.
The quality of the various analytical approximation is illustrated in Fig.~\ref{fig:q}. The exact function is 
contrasted with: (i) the plain 8.5PN-accurate function as calculated in Ref.~\cite{Bini:2014nfa}; 
(ii) its resummed version without the GSF-informed correction factor; (iii) the full, GSF-informed, 
function obtained from Eq.~\eqref{eq:hatq}. The bottom panel of the figure displays the relative
difference between the exact and the GSF-informed function.

\subsection{Radiation reaction and waveform}
\label{sec:fluxes}
Now that we have discussed in detail the structure of the conservative part of the model, let us
turn to remind the elements of the waveform and of the radiation reaction.
First of all, let us fix our waveform convention. The waveform strain is decomposed in
spin-weighted spherical harmonics as
\be
h_+ - i h_\times = \dfrac{1}{D_L}\sum_\ell \sum_{m=-\ell}^{\ell}h_\lm{}_{-2}Y_\lm(\iota,\phi),
\ee
where $D_L$ indicates the luminosity distance, and ${}_{-2}Y_\lm(\iota,\phi)$ are the $s=-2$
spin-weighted spherical harmonics, $\iota$ is the inclination angle with respect to the orbital
plane and $\phi$ the azimuthal one. 

The form of the radiation reaction, that has either an azimuthal, ${\cal F}_\varphi$, and a radial ${\cal F}_r$
component is precisely the one used in Ref.~\cite{Nagar:2021gss,Nagar:2021xnh} and so far tested in many
context, especially in the test-mass limit~\cite{Albanesi:2021rby,Albanesi:2022ywx}.
For the waveform, we consider all multipoles up to $\l=8$, though excluding the $m=0$ ones, that have
not yet been suitably validated within the EOB approach. All modes up to $\ell=m=5$ implement the 
general, Newtonian prefactor so to effectively incorporate many high-order corrections that are 
important along eccentric orbits~\cite{Chiaramello:2020ehz}. This requires the numerical calculation 
of several high-order time derivatives to compute the Newtonian prefactors, that can be 
explicitly found in Ref.~\cite{Albanesi:2021rby}. For the $\l=m=2$ mode we also consider the 
2PN noncircular corrections, both in the instantaneous part~\cite{Albanesi:2022xge} and in 
the hereditary contribution~\cite{Placidi:2021rkh}.

\section{An illustrative example}
\label{sec:example}
\begin{figure}[t]
	\includegraphics[width=0.45\textwidth]{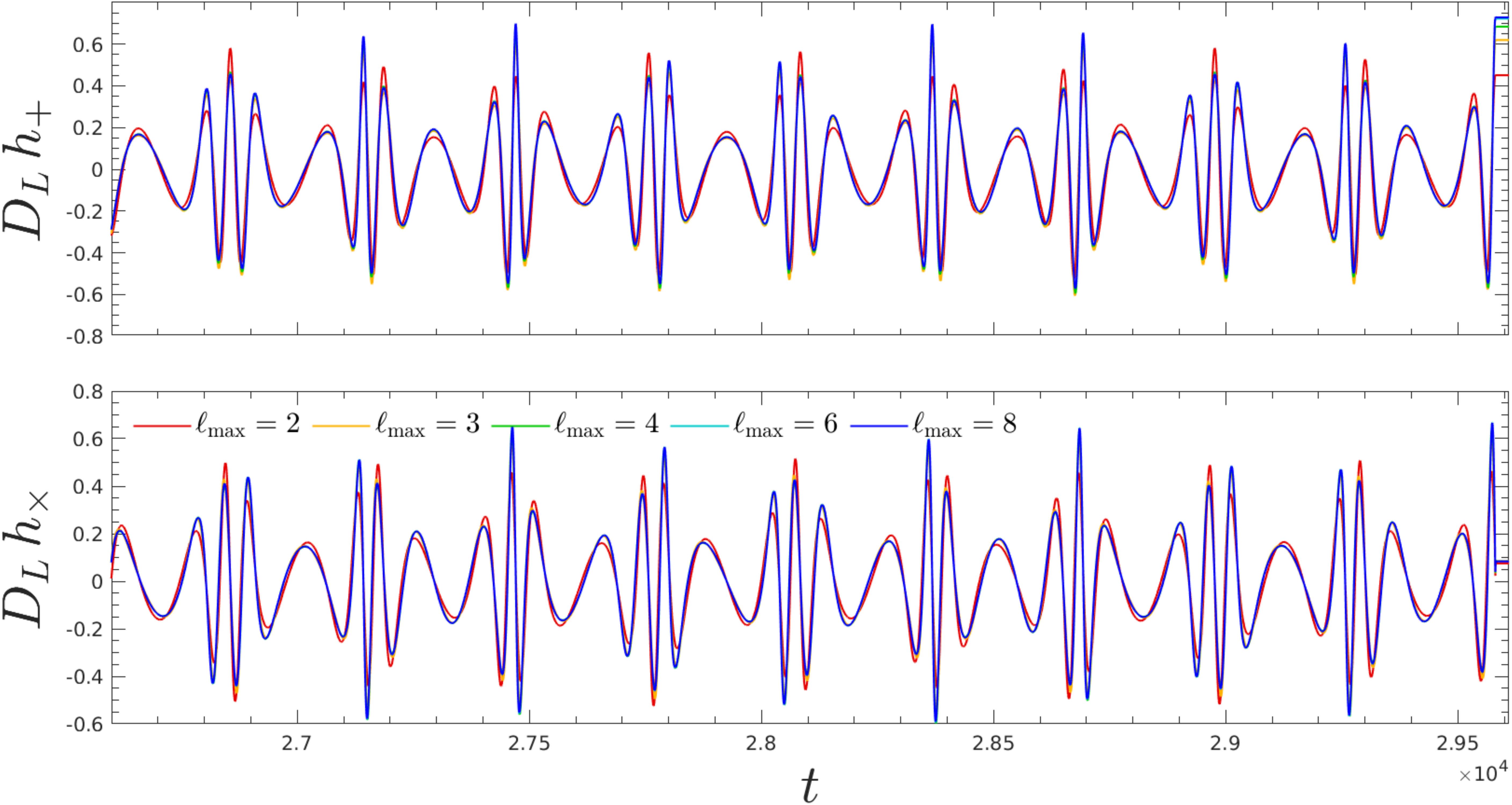} \\
	\includegraphics[width=0.22\textwidth]{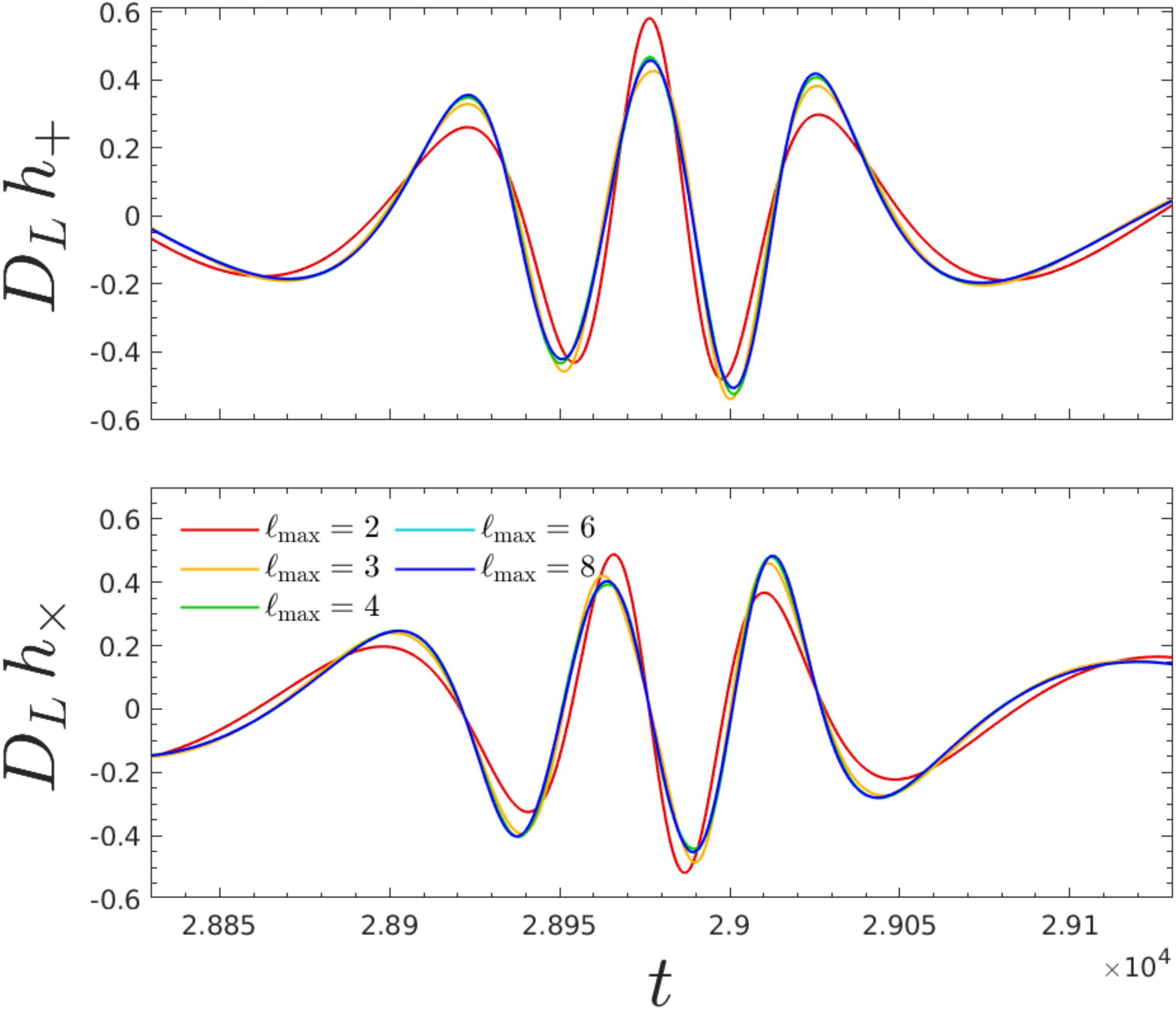}
	\hspace{0.08cm}
	\includegraphics[width=0.20\textwidth]{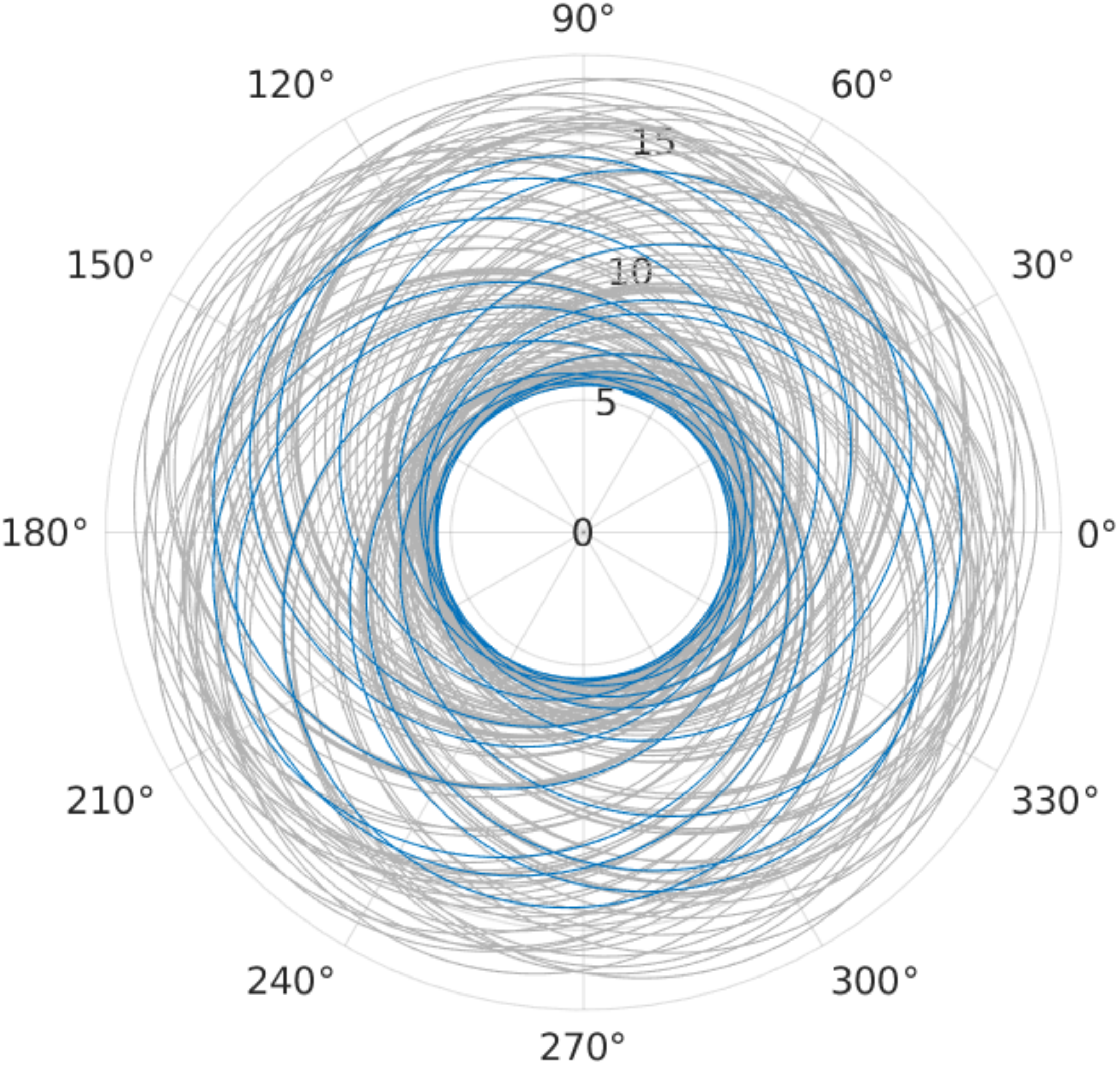}
	\caption{\label{fig:gsf_q100_hpc} 
		Strain generated by a binary with $q=10^3$, $e_0=0.5$, $\chi_1 = 0.3$, and $\chi_2=0.1$,
		as seen by an observer whose line of sight is inclined by $45^\circ$ with the orbital plane.
		In order to highlight the relevance of the higher modes, we show the strain 
		computed using different values for $\l_{\rm max}$. We do not include $m=0$ modes. 
		We also show a zoom on a radial period (left bottom panel).
		The portion of the strain shown in the upper panel corresponds to the trajectory highlighted
		in blue in the right bottom panel.
	}
\end{figure}
Let us finally conclude with an illustrative example waveform that can be generated with our GSF-informed formalism.
We consider an illustrative $q=1000$ binary\footnote{Evidently, there are no theoretical limitations in 
choosing an even larger mass ratio, either in the EMRI or  IMRI regime, but the computational cost 
would increase. A precise assessment of the computational cost of an actual $10^5$ EMRIs with 
the GSF-informed potentials is postponed to future work, still see Appendix~C of 
Ref.~\cite{Nagar:2021gss} for a general idea with the standard potentials}  with dimensionless spins 
$\chi_1\equiv S_1/m_1^2=0.3$, $\chi_2\equiv S_2/m_2^2=0.1$, initial EOB  eccentricity 
$e_0=0.5$ and initial semilatus rectum $p_0\simeq 8.68$. For the Newtonian-like 
definition of the (gauge-dependent) eccentricity and  semilatus rectum 
see e.g. Ref~\cite{Hinderer:2017jcs,Chiaramello:2020ehz,Nagar:2021gss,Albanesi:2021rby}. 
The trajectory and the corresponding strain computed considering all the $m>0$ modes
up to different values of $\l_{\rm max}$ are shown in Fig.~\ref{fig:gsf_q100_hpc},
where we consider an observer whose line of sight is inclined by $45^\circ$ with
respect to the orbital plane. Note that the evolution stops at $t\simeq 2.96\times 10^4 M$, 
slightly before the crossing of the EOB light-ring, but we only show the waveform 
that corresponds to the last few orbits. 
As can be clearly seen, especially in the bottom-left panels, the $\l=2$ modes 
alone are not sufficient to accurately catch all the waveform features, as expected a priori.
\begin{figure}[t]
	\includegraphics[width=0.45\textwidth]{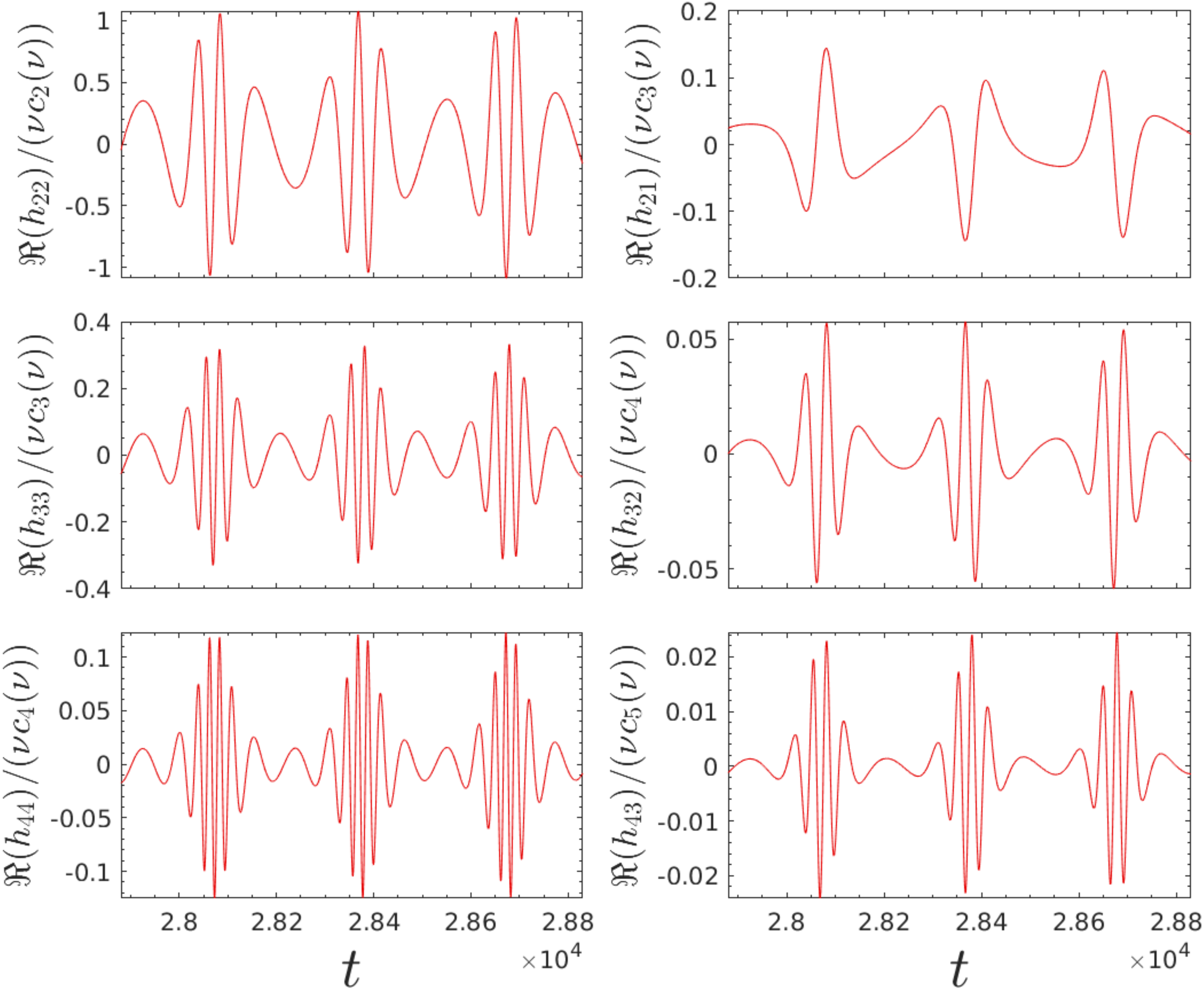} 
	\caption{\label{fig:gsf_q100_hlm} 
		Waveform multipoles for the same configuration considered in Fig.~\ref{fig:gsf_q100_hpc}.
		We show the $(2,2), (2,1), (3,3), (3,2), (4,4),$ and $(4,3)$ modes.}
\end{figure}
Some of the waveform multipoles are shown in Fig.~\ref{fig:gsf_q100_hlm}.
Each waveform multipole is normalized by the leading-order $\nu$-dependence $\nu c_{\ell +\epsilon}(\nu)$, 
where $\epsilon=\pi(\ell +m)$  is the parity of $\ell+m$, $\epsilon=0$ if $\ell+m$ is even 
and $\epsilon=1$ if $\ell+m$ is odd. The coefficients $c_{\ell +\epsilon}$ are defined as
\be
c_{\ell +\epsilon}(\nu)=X_2^{\ell +\epsilon-1}+(-)^m X_1^{\ell +\epsilon-1},
\ee
where $X_i\equiv m_i/M$. Note that the Newtonian noncircular corrections introduced 
in Ref.~\cite{Chiaramello:2020ehz} are included in all the modes up to $\l \leq 4$ and 
in the $\l=m=5$ mode. 

\begin{figure}[t]
	\includegraphics[width=0.45\textwidth]{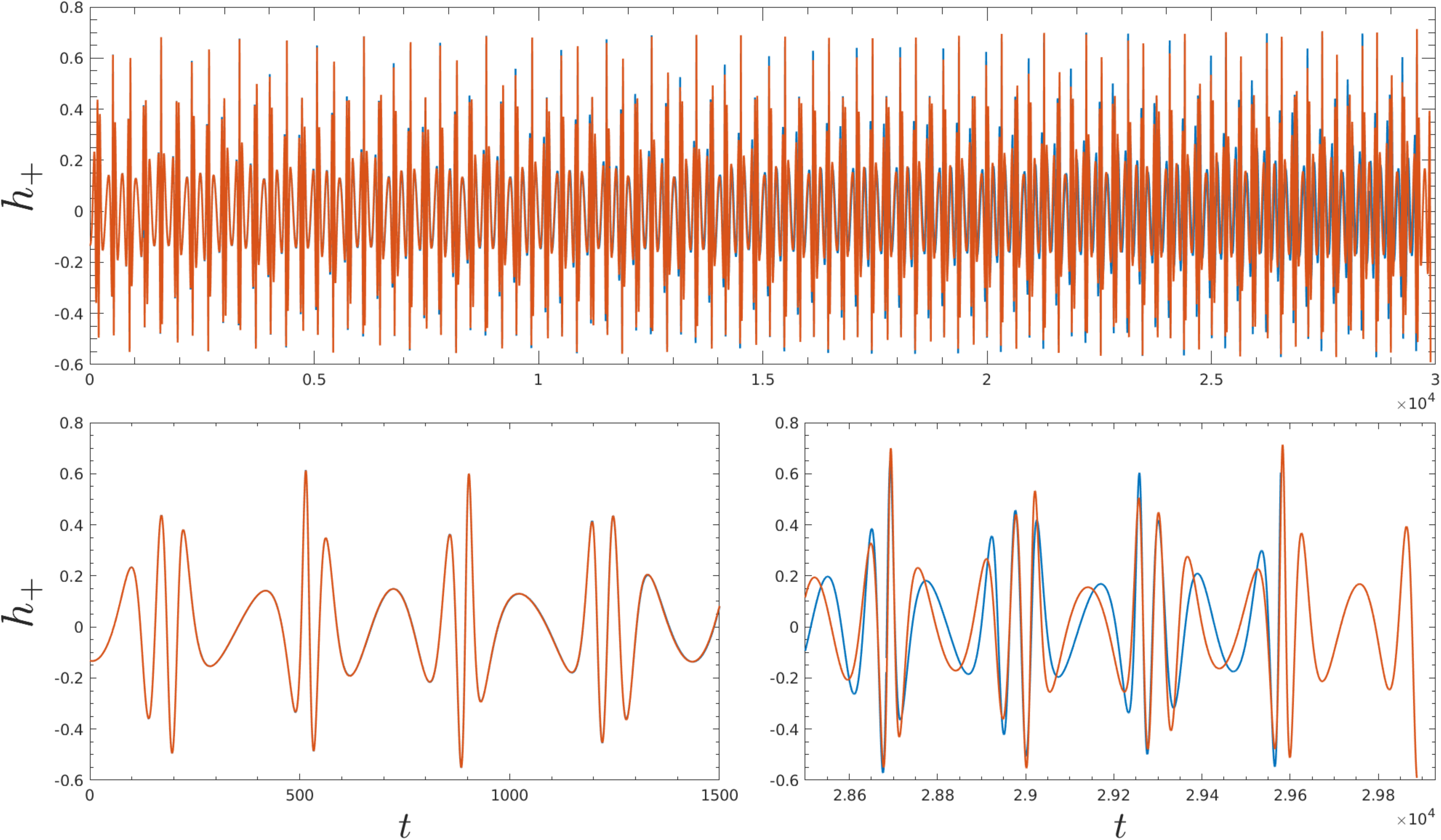} 
	\caption{\label{fig:eob_vs_kerr}
	Same initial data considered in Fig.~\ref{fig:gsf_q100_hpc} and Fig.~\ref{fig:gsf_q100_hlm}. 
	In the upper panel we show two $h_+$'s: the first one(blue) corresponds to the dynamics computed with GSF-potentials,
	while the other one (orange) is obtained from the dynamics computed with Kerr potentials, i.e. without $\nu$-dependent
	corrections in the Hamiltonian. The bottom panels close up on the beginning of the inspiral and on the end.
	We consider all $m\neq 0$ multipoles up to $\l=8$.
	}
\end{figure}

In order to assess the relevance of the GSF $\nu$-corrections, 
we compare the evolution discussed above with the one obtained using the same
initial data but setting to zero the $\nu$ corrections in $(A,\bar{D},Q)$, 
i.e. using the Kerr metric functions. The corresponding waveforms can be found 
in Fig.~\ref{fig:eob_vs_kerr}, where we also highlight the first and the last few orbits. 
As can be seen, the conservative GSF $\nu$-corrections are not negligible since their
absence leads to a clear dephasing that translates in a longer evolution.

Finally, it should be taken into account that also the non-conservative dynamics 
is crucial for the evolution of EMRIs and IMRIs. The accuracy of the fluxes 
(and thus of the radiation reaction) for noncircular orbits has been 
already discussed in Refs.~\cite{Albanesi:2021rby,Albanesi:2022ywx},
but further improvements are needed in order to describe real astrophysical
scenarios. For this reason, in Appendix~\ref{appendix:qc_flux} 
we discuss how the flux can be improved in the simplest case 
of (quasi)-circular orbits in Schwarzschild spacetime. 

\section{Conclusions and future directions}
\label{sec:end}
\begin{figure}[t]
	\vspace{0.4cm}
	\includegraphics[width=0.45\textwidth]{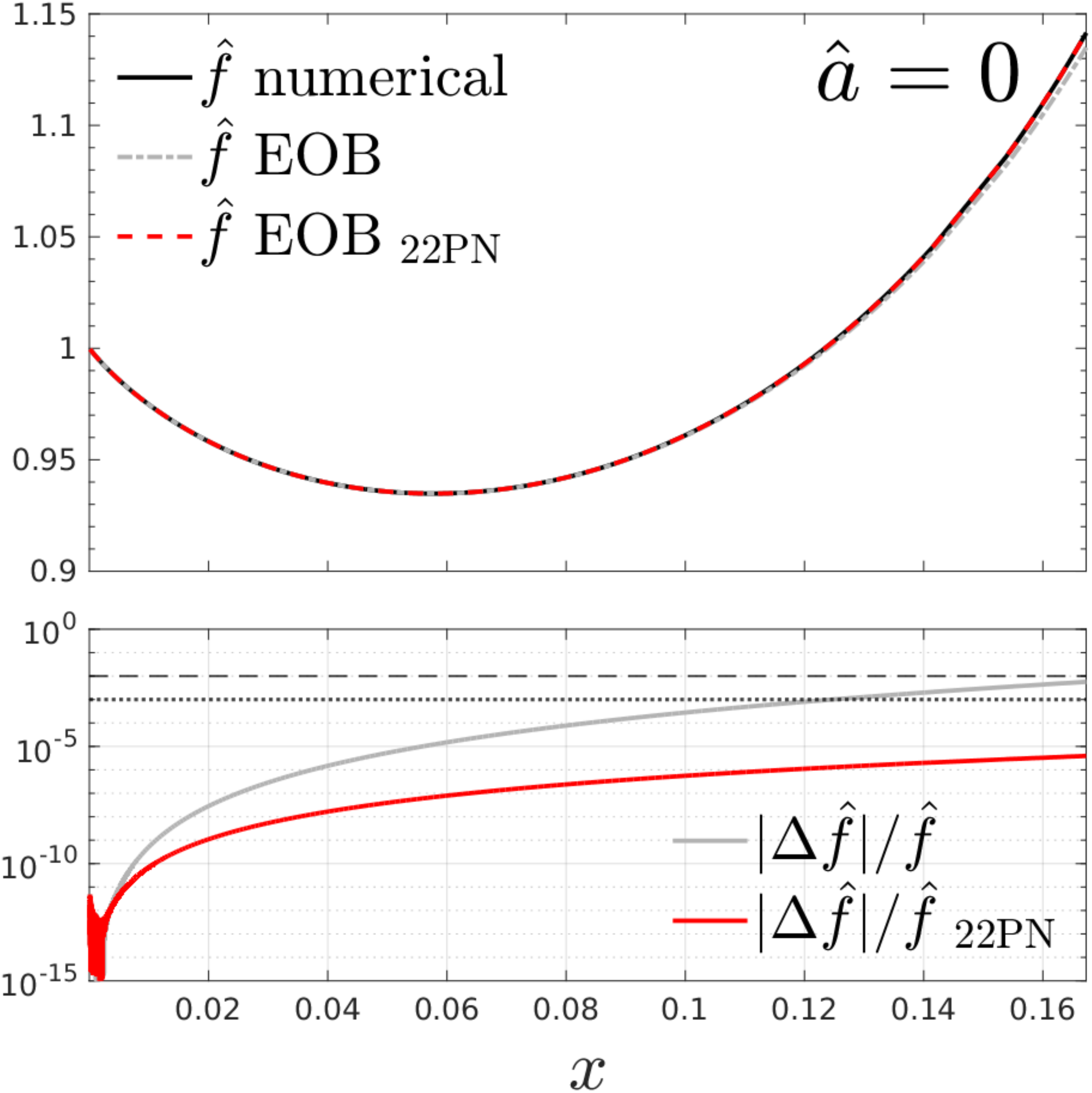} 
	\caption{\label{fig:testmass_hatf}
		Numerical (black) and EOB (gray and red) fluxes in Schwarzschild plotted 
		against $x=\Omega^{2/3}$ up to the LSO. We consider all the $m>0$ modes up to
		$\ell=8$. The flux in gray is computed using
		the standard flux implemented in \TEOBResumSDali{}, while the red one 
		is computed similarly but employing 22PN information in the residual 
		amplitudes $\rho_\lm(x)$.
	}
\end{figure}
Building upon the eccentric version of the \TEOBResumS{} model~\cite{Nagar:2021xnh,Gamba:2021ydi}, 
the {\tt TEOBResumS-DALI} model~\cite{Nagar:2021gss,Nagar:2021xnh},
we have constructed the first GSF-informed EOB waveform model for spin-aligned, eccentric binaries. This model crucially 
incorporates GSF-informed correction in the orbital sector and should be seen as a first step towards the construction
of a physically complete model for IMRI and EMRIs. As such, it does not incorporate the transition from the inspiral
to plunge, merger and ringdown. Our main finding is to have shown that suitable resummation strategies allow us to 
improve the strong-field behavior of (high-order) PN expansion of the EOB potentials $(A,\bar{D},Q)$. This gives consistency
between the various PN orders, with a qualitative and semi-quantitative good agreement with the
exact potentials computed numerically. We note that this resummation is actually {\it different} from the one routinely 
applied to the NR-informed potentials of \TEOBResumS{} valid in the comparable-mass case.
Several technical improvements are  however necessary in order to reach the goal of an analytically complete, and faithful, 
EOB-based waveform model for IMRIs and EMRIs. Let us list here a few:
\begin{enumerate}
\item[(i)]{} In our model, the analytic resummations of the EOB functions $(A,\bar{D},Q)$ are improved by effective, high-order,
             corrections that are informed by fitting to GSF numerical data. The accuracy of the procedure depends on
             both the initial (resummed) PN order of $(A,\bar{D},Q)$ and the functional form of the correcting 
             factor\footnote{Evidently, this also depends on the accuracy of the original GSF numerical data, 
             that might need improvements especially between the LSO and the light-ring~\cite{Akcay:2015pjz}.}.
             The choices made here might be improved if needed, and this would require a dedicated study on its
             own.
\item[(ii)]{} For simplicity, here we have decided to use the standard \TEOBResumS{} spin sector (although {\it without} the NR-tuning),
              that implements next-to-next-to-leading order spin-orbit accuracy~\cite{Nagar:2011fx,Damour:2014sva}. This
              accuracy is {\it not} state of the art, since complete analytic information (beyond the linear-in-$\nu$ knowledge) 
              to the next-to-leading order is available~\cite{Antonelli:2020ybz,Antonelli:2020aeb}.
              In addition, one can also incorporate the full 1GSF information in the spin-orbit coupling functions computed
              in several works~\cite{Bini:2015xua,Kavanagh:2017wot} and informed to exact GSF data.
              At a more radical level, however, the spin-orbit structure of \TEOBResumS{} should be modified so to incorporate
              the {\it complete} leading-order contribution of the spinning secondary~\cite{Barausse:2009aa}, 
              that it is now only approximated. A route to do so within the \TEOBResumS{} framework was 
              suggested in the Conclusions of Ref.~\cite{Rettegno:2019tzh}. Note also that this approach could
              be pushed to one further PN order using the analytical results of Refs.~\cite{Antonelli:2020ybz,Antonelli:2020aeb}.
              Since the current work is mainly illustrative of the potentialities of a GSF-informed EOB framework
              for IMRIs and EMRIs, we postpone this detailed study to future work.
\item[(iii)]{}Similarly, for simplicity here we are using the fluxes of \TEOBResumS{}, that, in the quasi-circular limit, rely
              on $3^{+2}$PN and $3^{+3}$PN in various multipoles (up to $\ell=8$) that are then resummed with various combinations
              of Pad\'e approximants~\cite{Nagar:2016ayt, Messina:2018ghh,Nagar:2020pcj}
              In principle, the analytical accuracy of the test-mass contribution to the fluxes should be improved to describe 
              more accurately the radiation-reaction driven long-inspiral of large mass ratios binaries. This was already pointed
              out in Ref.~\cite{Nagar:2022icd}, thanks to a specific EOB/NR comparison involving the standard \TEOBResumS{} and
              a recently obtained $q=15$ NR simulation~\cite{Yoo:2022erv}. The inclusion of more test-mass terms in the resummed 
              flux (notably up to 22PN accuracy~\cite{Fujita:2012cm}) brings a much closer agreement between analytical and numerical 
              fluxes in the test-mass limit. This is discussed in some detail in Appendix~\ref{appendix:qc_flux} and will be taken
              in due account in future work.
\item[(iv)]{}The waveform model discussed here should be validate against full, self consistent, GSF evolutions~\cite{Wardell:2021fyy}. 
                  This will allow us to understand the impact of the approximations we used (e.g. the one for the flux) in the context of
                  actual EMRIs. This analysis will be tackled in forthcoming work.
\end{enumerate}
In conclusion, we have shown that, once informed by (sparse) GSF data, the current EOB formalism is suitable for constructing
waveforms for EMRIs (and IMRIs). The main open issues are in improving the accuracy of certain building blocks of the model
for LISA purposes (e.g. fluxes or spin sector), but the model is conceptually complete at it is clear how to improve it. Given this
model, it will be interesting to investigate to which extent standard twisting techniques (see e.g.~\cite{Akcay:2020qrj,Gamba:2021ydi} 
and references therein) usually employed for obtaining precessing waveforms are reliable in the context of EMRIs. It will be
similarly easy to include environmental effects on the dynamics, like those related to the presence of an accretion disk, 
in the same spirit of~\cite{Speri:2022}.
\begin{acknowledgments}
A.N. thanks to C.~Kavanagh for computing for us the 25PN accurate expressions of
the $A$ function and for collaboration at IHES at the very beginning of this work. 
Discussions with D.~Bini and R.~Gamba are also gratefully acknowledged. We 
also thank A.~Albertini and P.~Rettegno for help during the development of 
this work, and S.~Akcay and M.~van de Meent for additional information 
on their numerical data and for comments.
\end{acknowledgments}

\appendix
\section{Improving the (quasi)-circular nonspinning flux}
\label{appendix:qc_flux}
In this Appendix we briefly discuss how the radiation reaction force related
to the flux at infinity can be improved in the nonspinning case. The angular 
component of the radiation reaction $\hat{\F}_\varphi$ in \TEOBResumS{} is written factorizing 
the Newtonian quadrupolar contribution
\begin{equation}
	\hat{\F}_\varphi = - \frac{32}{5} \nu r_{\rm \omega}^4 \Omega^5 \hat{f},
\end{equation}
where $r_\omega$ is a suitably defined radius that satisfies a generalized Kepler's 
law~\cite{Damour:2014sva} and $\hat{f}$ is the reduced flux function, that contains 
all the high-PN corrections recasted via suitable factorization and 
resummations~\cite{Damour:2008gu}.
A crucial aspect to obtain very reliable fluxes is the accuracy of
the residual general relativistic PN corrections to the amplitude, denoted
with $\rho_\lm$, that enter in $\hat{f}$ at power $2\ell$~\cite{Damour:2008gu}. 
Suitable factorization and resummations for these corrections have been
discussed in Refs.~\cite{Nagar:2016ayt, Messina:2018ghh}.
The standard analytical resummed implementation of $\hat{f}$ used in 
\TEOBResumSDali{}, that relies on Pad\'e resummed $\rho_\lm$'s with 6PN accuracy
up to $\ell=6$, is shown in gray in Fig.~\ref{fig:testmass_hatf}. The curve is
compared with the exact data (black) obtained numerically using S.~Hughes code~\cite{Hughes:2005qb}
and already used as benchmark in previous  studies~\cite{Nagar:2016ayt,Messina:2018ghh,Albanesi:2021rby,Albanesi:2022ywx}.
As can be seen, the analytical/numerical disagreement (gray) increases 
monotonically and almost reach the $0.6\%$ at the LSO. However, the reduced flux function 
can be immediately improved considering the Taylor-expanded
series for the $\rho_\lm$ up to 22PN~\cite{Fujita:2012cm}. The flux obtained in this way 
is also shown in Fig.~\ref{fig:testmass_hatf} (red) and it is systematically more accurate
than the standard \TEOBResumS{} flux. Even at the LSO, the 
analytical/numerical disagreement remains below the $4\times10^{-6}$ level.

\bibliography{refs20220728.bib,local.bib}

\end{document}